# Fast Computation of Isochrones in Road Networks

Moritz Baum[1], Valentin Buchhold[1], Julian Dibbelt[1] and Dorothea Wagner[1]

[1]Karlsruhe Institute of Technology, Germany, `first.last@kit.edu`

November 12, 2015

### Abstract

We study the problem of computing isochrones in road networks, where the objective is to identify the region that is reachable from a given source within a certain amount of time. While there is a wide range of practical applications for this problem (e. g., reachability analyses, geomarketing, visualizing the cruising range of a vehicle), there has been little research on fast computation of isochrones on large, realistic inputs. In this work, we formalize the notion of isochrones in road networks and present a basic approach for the resulting problem based on Dijkstra's algorithm. Moreover, we consider several *speedup techniques* that are based on previous approaches for one-to-many shortest path computation (or similar scenarios). In contrast to such related problems, the set of targets is not part of the input when computing isochrones. We extend known Multilevel Dijkstra techniques (such as CRP) to the isochrone scenario, adapting a previous technique called isoGRASP to our problem setting (thereby, enabling faster queries). Moreover, we introduce a family of algorithms based on (single-level) graph partitions, following different strategies to exploit the efficient access patterns of PHAST, a well-known approach towards one-to-all queries. Our experimental study reveals that all speedup techniques allow fast isochrone computation on input graphs at continental scale, while providing different tradeoffs between preprocessing effort, space consumption, and query performance. Finally, we demonstrate that all techniques scale well when run in parallel, decreasing query times to a few milliseconds (orders of magnitude faster than the basic approach) and enabling even interactive applications.

## 1. Introduction

Web-based map services, autonomous navigation systems, and other location-based applications have gained wide currency in the last decades, motivating a great deal of research on practical algorithms for routing in road networks [2]. Most work focused on computing distances (or shortest paths) between pairs (or sets) of vertices. On the other hand, *isochrones* (more generally, *isolines*) are defined as curves of constant distance (from a given source). Thereby, isochrones represent the area that is in range for a given time limit (or some other limited resource). Hence, rather than actual distances, information about the subgraph within range is required. Isolines are relevant in a wide range of applications, such as reachability analyses [3, 17, 18], geomarketing [13], range visualization for (electric) vehicles [20], and a variety of online applications [22].

**Related Work.** Dijkstra's well-known algorithm [12] computes shortest paths from a given source to all (reachable) vertices in (almost) linear time. The MINE algorithm [17] is a Dijkstra-based search to compute isochrones in transportation networks, using a spatial network database. An improved variant, MIMEX [18], has reduced space requirements. Both approaches work on databases, prohibiting interactive applications (with running times beyond the order of minutes).

Regarding shortest-path computation, on the other hand, various *speedup techniques* [2] were developed for faster queries, using an offline preprocessing phase (sometimes including an additional *customization* step to incorporate, e. g., user preferences or traffic updates), and an online query phase. Customizable Route Planning (CRP) [5] uses overlays based on a (multilevel) graph partition to speed up queries [8, 21, 23]. Metric customization is practical, as only shortcut costs in the overlay need to be updated. In Contraction Hierarchies (CH) [19], vertices are contracted in increasing order of importance, creating shortcuts between neighbors to maintain distances. A customizable variant of CH (CCH) was introduced in [11].

Both CH and CRP were extended to related scenarios, such as *one-to-many* or *one-to-all* queries. Delling et al. introduce PHAST [4], exploiting CH for fast one-to-all and all-pairs shortest-path computations. Queries consist of a CH search and a linear sweep over the vertices to obtain distances. For one-to-many scenarios, RPHAST [7] uses a *target selection* phase to speed up the second phase of PHAST queries. Efentakis et al. [14] propose GRASP, a one-to-all (and one-to-many) technique that builds upon CRP, enabling customization in a few seconds. *Many-to-many* queries can be handled by bucket-based approaches [24], storing pairs of target and distance at vertices. This generic approach can be implemented by several speedup techniques [7]. Other related query types include point-of-interest (POI) queries, best-via queries (shortest paths that visit certain types of POI), and $k$-nearest neighbors (kNN) queries. Bucket-based approaches can be extended to such scenarios [1]. Delling et al. [9] support POI and kNN queries based on CRP. Efentakis et al. [15] extend GRASP to cover kNN queries. All techniques exploit the fact that targets (or POIs) are known in advance. To the best of our knowledge, the only speedup technique extended to isochrone queries is GRASP [14].[1] However, GRASP computes distances to all vertices in range, which can be wasteful if only the actual isoline is required.

**Our Contribution.** This work studies speedup techniques for isochrone computation. Since no canonical definition exists in the literature, we formally define the isochrone problem (specifying the output of our algorithms) and present a variant of Dijkstra's algorithm to solve it. For faster queries, we propose a new variant of the CRP speedup technique that computes isochrones. As an alternative, we also extend the existing approach of isoGRASP [14] to our definition. Next, we introduce a family of novel approaches that exploit different strategies to combine graph partitions and (R)PHAST. We also show how all approaches can be parallelized for further speedup. Our experimental evaluation on realistic, large-scale input reveals that all techniques are orders of magnitude faster than the basic approach. Providing queries in the order of milliseconds, they enable a range of new (e. g., interactive) applications. Since all proposed techniques differ in terms of customization effort, memory consumption, and query time, each is suitable for certain applications.

**Outline.** Section 2 introduces basic terminology and building blocks. Section 3 formally defines the problem and provides a basic approach. Section 4 presents techniques based on CRP and GRASP, while Section 5 introduces algorithms that combine graph partitions and PHAST. Section 6 experimentally evaluates all approaches. Section 7 concludes with final remarks.

## 2. Preliminaries

We consider road networks given as *directed graph* $G = (V, E)$ with *length function* len: $E \to \mathbb{R}_{\geq 0}$ (representing, e. g., travel time). An *s–t-path* in $G$ is a sequence $P_{s,t} = [s = v_1, v_2 \ldots, v_k = t]$ of vertices, such that $(v_i, v_{i+1}) \in E$ for $1 \leq i \leq k-1$. The length of $P_{s,t}$ is the sum of its edge lengths. We presume that graphs are *strongly connected*, i. e., there exists an *s–t*-path for each pair $s, t \in V$. The *distance* $d(s, t)$ from $s$ to $t$ is the length of the shortest *s–t*-path in $G$.

A *(vertex) partition* is a family $\mathcal{V} = \{V_1, \ldots, V_k\}$ of *cells* $V_i \subseteq V$, such that $V_i \cap V_j = \emptyset$ for $i \neq j$ and $\bigcup_{i=1}^{k} V_i = V$. A *(nested) multilevel partition* with $L$ levels is a family $\Pi = \{\mathcal{V}^1, \ldots, \mathcal{V}^L\}$ of partitions of nested cells, i. e., for each level $\ell \leq L$ and cell $V_i^\ell \in \mathcal{V}^\ell$, there is a cell $V_j^{\ell+1} \in \mathcal{V}^{\ell+1}$ at level $\ell+1$ with $V_i^\ell \subseteq V_j^{\ell+1}$. For consistency, we define $\mathcal{V}^0 := \{\{v\} \mid v \in V\}$ (the trivial partition where each vertex has its own cell) and $\mathcal{V}^{L+1} := \{V\}$ (the trivial single-cell partition). An edge $(u, v) \in E$ is a *boundary edge* ($u$ and $v$ are *boundary vertices*) on level $\ell$, if $u$ and $v$ are in different cells of $\mathcal{V}^\ell$. Similar to vertex partitions, we define *edge partitions* $\mathcal{E} = \{E_1, \ldots, E_k\}$, with $E_i \cap E_j = \emptyset$ for $i \neq j$ and $\bigcup_{i=1}^{k} E_i = E$. A vertex $v \in V$ is *distinct* (wrt. $\mathcal{E}$) if all its incident edges belong to the same cell, else $v$ is a *boundary vertex* or *ambiguous*.

---
[1] An extension of the CRP approach in [9] to isochrones is outlined in a patent description (US Patent App. 13/649,114; http://www.google.com/patents/US20140107921), however, in a simpler than our intended scenario. Furthermore, the approach was neither implemented nor evaluated.

Dijkstra's algorithm [12] computes, for a given source $s$, the distances $d(s, v)$ to all $v \in V$. It maintains *distance labels* $d(\cdot)$ for each vertex (initially, $d(s) = 0$ and $d(v) = \infty$ for $v \neq s$). In each iteration, the algorithm extracts a vertex $u$ with minimum $d(u)$ from a priority queue (initialized with $s$) and *settles* it. At this point, $d(u)$ is *final*, i.e., $d(u) = d(s, u)$. It then *scans* all edges $(u, v)$, i.e., if $d(u) + \text{len}(u, v) < d(v)$, it updates $d(v)$ accordingly and adds (or updates) $v$ in the queue.

**Speedup Techniques.** The three-phase workflow of CRP [5] distinguishes preprocessing and metric customization. First, a (multilevel) partition $\Pi$ of the road network is computed, inducing for each level $\ell$ of $\Pi$ an *overlay graph* $H^\ell$ that contains all boundary vertices and boundary edges in $\mathcal{V}^\ell$ and *shortcut* edges between boundary vertices of each cell $V_i^\ell \in \mathcal{V}^\ell$. During customization, the lengths of all shortcuts are computed (using, e.g., Dijkstra's algorithm). For fast integration of new length functions, previously computed overlays are used to compute shortcuts of higher levels. For $s$–$t$-queries, Dijkstra's algorithm is run on the union of the top-level overlay $H^L$ and the subgraphs (of $H^\ell$ for $0 \leq \ell \leq L-1$, $H^0 := G$) induced by the cells containing $s$ or $t$.

In CH [19], vertices are contracted in a (heuristic) order $\text{rank}: V \to \{1, \ldots, |V|\}$. To contract a vertex $v$, shortcut edges are added between uncontracted neighbors (if necessary) to preserve distances. Also, *levels* $\ell(\cdot)$ are assigned to vertices, initially set to zero. When contracting $u \in V$, we set $\ell(v) = \max\{\ell(v), \ell(u) + 1\}$ for each uncontracted neighbor $v$. Let $E^+$ denote the set of shortcuts added during preprocessing. An $s$–$t$-query is bidirectional, with a forward search (from $s$) on $G^\uparrow = (V, E^\uparrow)$, where $E^\uparrow = \{(u, v) \in E \cup E^+ : \text{rank}(u) < \text{rank}(v)\}$, and a backward search (from $t$) on $G^\downarrow = (V, E^\downarrow)$, where $E^\downarrow = \{(u, v) \in E \cup E^+ : \text{rank}(u) > \text{rank}(v)\}$.

**Batched Shortest Paths.** Both CH and CRP compute distances between pairs of vertices. GRASP [14] extends CRP to batched query scenarios. In addition to shortcuts between boundary vertices (within a cell), each level-$\ell$ boundary vertex (for $0 \leq \ell < L$) stores (incoming) *downward shortcuts* from boundary vertices of its corresponding supercell at level $\ell + 1$. Customization works similar to CRP, storing downward edges in a separate *downward graph* $H^\downarrow$. For one-to-all queries, the *upward phase* runs a CRP search from the source, obtaining distances to settled vertices. Then, the *scanning phase* processes cells in descending level order, sweeping over downward edges to propagate distance labels from boundary vertices to those at the level below.

PHAST [4] exploits CH preprocessing for fast one-to-all queries. The *upward phase* runs a forward CH search, while the *scanning phase* processes vertices in descending order of CH level and propagates distances by scanning incoming edges in $E^\downarrow$. Vertices are reordered during preprocessing, so the scanning phase is a linear sweep over an edge array.

RPHAST [7] enables one-to-many queries by introducing an additional *target selection* phase. Given a target set $T$, it extracts a (restricted) subgraph $G_T^\downarrow$ of the original downward graph with a BFS in $G^\downarrow$ (from all $t \in T$). Queries resemble PHAST queries, running the scanning phase on $G_T^\downarrow$ instead of $G^\downarrow$.

## 3. Problem Statement and Basic Approach

We formalize the notion of isochrones and describe a basic algorithm. Given a graph $G = (V, E)$ and a length function len on its edges, the *isochrone problem* takes as input a source $s \in V$ and a time limit $\tau \in \mathbb{R}_{\geq 0}$. We say that a vertex $v \in V$ is *in range* if $d(s, v) \leq \tau$, else it is *out of range*. There is no canonical definition of the output of the isochrone problem. Usually, a compact representation of all vertices in range (or edges with at least one incident vertex in range) is required [14, 18, 25]. Therefore, we define the output of the isochrone problem as the set of all *isochrone edges* that separate vertices in range from those out of range. Observe that these are the edges $(u, v) \in E$ with exactly one endpoint in range, i.e., either $d(s, u) \leq \tau, d(s, v) > \tau$ or $d(s, u) > \tau, d(s, v) \leq \tau$. This set of edges compactly represents the area in range (thus, generating output is unlikely to become a performance bottleneck). All approaches presented below can be modified to serve other definitions (requiring, e.g., the set of vertices in range).

Next, we describe a variant of Dijkstra's algorithm, called *isoDijkstra*, to compute all isochrone edges for a given source $s$ and time limit $\tau$. The search runs from $s$ as described in Section 2, but stops once the *stopping criterion* is fulfilled, i.e., the distance label of the minimum element in the queue exceeds $\tau$. To determine isochrone edges, we then sweep over all vertices $v$ still contained in the queue (note that these must be out of range). For each, we check all incoming edges $(u, v)$ and output exactly those where $u$ is in range. Clearly, this way we find all isochrone edges $(u, v)$ where $u$ is in range (Dijkstra's algorithm scans exactly the vertices in range, so the out-of-range endpoints of all isochrone edges were added to the queue, but not extracted). We make the following modification to ensure that isochrone edges $(u, v)$ where $u$ is out of range are found as well. When settling a vertex $u$, we also scan incoming edges $(v, u)$. If $d(v) = \infty$, we insert $v$ into the queue with a key of infinity. Thereby, we guarantee that both types of isochrone edges are contained in the queue when the search terminates.

## 4. Multilevel Dijkstra Approaches

The basic idea for Multilevel Dijkstra (MLD) approaches is to skip cells that are in range, but descend into lower levels where necessary to determine isochrone edges. Following this approach, we have to cope with several challenges. First, it is not sufficient to check whether the time limit is exceeded at boundary vertices (we may miss isochrone edges that are part of no shortcut, see Appendix A for an example). Therefore, we compute additional data during customization. Second, when descending into lower levels, distance labels must be consistent within each cell. This motivates a two-phase approach. We say that a cell is *active*, if its induced subgraph contains at least one isochrone edge. The first phase determines active cells (running a CRP query), while the second phase descends into lower-level overlays to determine isochrone edges.

**Customization.** Metric customization works along the lines of plain CRP. We compute distances of clique shortcuts by running Dijkstra searches (restricted to the respective cells). Cliques are represented as square matrices in contiguous memory for efficiency. To improve data locality and simplify index mapping, vertices are reordered such that boundary vertices are pushed to the front (ordered by descending level), breaking ties by cell [5].

Given a boundary vertex $u$ of a cell $V_i^\ell$, we define its *(level-$\ell$) eccentricity*, denoted $\text{ecc}_\ell(u)$, as the maximum (finite) distance $d(u,v)$ to some $v \in V_i^\ell$ on the subgraph induced by $V_i^\ell$. Note that this subgraph is not strongly connected in general, but restricting eccentricites to cells allows faster customization (see Appendix B for alternative approaches). We compute eccentricities for all boundary vertices on each level (storing them adds a single column to each clique matrix). To this end, we make use of the Dijkstra searches that compute shortcut lengths. Given a boundary vertex $v$ at level 1, we run the search from $v$ until the queue is empty (instead of aborting it once all boundary vertices were settled) at negligible overhead. The distance label of the last settled vertex is the eccentricity $\text{ecc}_1(v)$ of $v$. At higher levels ($1 < \ell \leq L$), we compute *upper bounds* on eccentricities as follows. When settling a vertex $v$, we check whether the sum of the label $d(v)$ and $\text{ecc}_{\ell-1}(v)$ exceeds the current bound and update it if necessary. While upper bounds may lead to cells falsely being marked active, this does not violate correctness of queries.

**isoCRP.** Given a source $s \in V$ and a limit $\tau$, queries run in two phases. First, the *upward phase* phase is similar to a plain CRP query from $s$, but running isoDijkstra on the search graph (the union of the top-level overlay and all induced subgraphs of cells containing $s$). Thus, the upward phase ends when all elements in the queue have a distance label greater than $\tau$. Additionally, to determine active cells, we maintain two flags $\texttt{i}(\cdot)$ (initially false) and $\texttt{o}(\cdot)$ (initially true) for each cell (per level), to indicate whether the the cell contains at least one vertex that is *in* or *out* of range, respectively. When settling a vertex $u \in V_i^\ell$, we set $\texttt{i}(V_i^\ell)$ to true if $d(u) \leq \tau$. Next, we check whether $d(u) + \text{ecc}_\ell(u) \leq \tau$. Observe that this condition is not sufficient to unset $\texttt{o}(V_i^\ell)$, because $\text{ecc}_\ell(u)$ was computed on the subgraph of $V_i^\ell$. If this subgraph is not strongly connected, $d(u) + \text{ecc}_\ell(u)$ is (in general) not an upper bound on the distance to any vertex in $V_i^\ell$ (there may be shorter paths to internal vertices of $V_i^\ell$ via boundary vertices that are unreachable from $u$ in the cell-induced subgraph). Therefore, when scanning an outgoing shortcut $(u, v)$ with length $\infty$ (this shortcut exists due to the matrix representation), we also check whether $d(v) + \text{ecc}_\ell(v) \leq \tau$. If the condition holds for $u$ and all boundary vertices $v$ unreachable from $u$ (wrt. $V_i^\ell$), we can safely unset $\texttt{o}(V_i^\ell)$. Note that all checks are cache efficient. Toggled flags are *final*, so we no longer need to perform any checks for such flags. After the upward phase finished, we mark all cells $V_i^\ell$ that have both $\texttt{i}(V_i^\ell)$ and $\texttt{o}(V_i^\ell)$ set as active (since isochrone edges can only be contained in cells having vertices both in and out of range).

The *downward phase* then consists of $L$ subphases. In decreasing order of levels, and for every active cell at the current level $\ell$, each subphase runs isoDijkstra restricted to the respective cell in $H_{\ell-1}$. Initially, all boundary vertices are inserted into the queue using their distance labels (according to the previous phase) as keys. Again, we check eccentricities on-the-fly to mark active cells for the next phase. Isochrone edges are determined at the end of each isoDijkstra search (see Section 3). On overlays, only boundary edges are reported (in contrast to shortcuts).

**isoGRASP.** We describe a modified version of isoGRASP [14] that is more suitable for our scenario (in [14], distances to *all* vertices in range are computed). Customization of our variant is similar to isoCRP, computing eccentricities of boundary vertices as described above. Additionally, we generate downward edges required by GRASP on-the-fly. We apply edge reduction (removing shortcuts via other boundary vertices) [14] to downward edges, but stick to the matrix representation for overlay edges. Queries again consist of two phases. The *upward phase* is identical to the one of isoCRP. The *scanning phase* handles levels from top to bottom in $L$ subphases and processes active cells. For an active cell $V_i^\ell$ at level $\ell$, isoGRASP sweeps over its *internal* vertices (all vertices in $H_{\ell-1}$ that lie in $V_i^\ell$ and are no boundary vertex of $V_i^\ell$). For each internal vertex $v$, its incoming downward edges are scanned, obtaining the distance at $v$. To determine active cells for the next subphase, we maintain flags $\texttt{i}(\cdot)$ and $\texttt{o}(\cdot)$ as described for isoCRP. This requires checks at unreachable boundary vertices (from $v$ within $V_i^\ell$). For speedup, we precompute these vertices and store them in a separate adjacency array.

Similar to isoCRP, the upward phase collects all isochrone edges that are boundary edges at level $L$ or contained in the cell-induced subgraph of the source. For remaining isochrone edges, we sweep over internal vertices a second time after processing a cell in the scanning phase. To avoid duplicates and to ensure that endpoints of scanned edges have correct distances, we skip edges leading to vertices with higher index (recall that we reorder vertices during preprocessing). Scanned boundary edges are then added to the output if exactly one endpoint is reachable.

**Parallelization.** During customization, cells of each level are processed in parallel [5]. To avoid concurrent memory access when customizing isoGRASP, we maintain thread-local containers for downward edges, which are concatenated at the end. Regarding queries, the (more expensive) downward phase is parallelized in a natural way, as cells at a certain level can be handled independently. Thus, we assign cells to threads and synchronize them between subphases. To reduce the risk of false sharing (i. e., concurrent access to the same cache line), we assign blocks of consecutive cells (wrt. vertex ordering) to the same thread. Moreover, to reduce synchronization overhead, we process cells on lower levels in a top-down fashion within the same thread.

## 5. Contraction-Based Approaches

This section introduces *isoPHAST*, an approach based on (R)PHAST to compute isochrones. Since the targets are not part of the input, we make use of graph partitions to restrict the area that is examined for isochrone edges. Queries have three phases, starting with a forward CH search from the source. Next, active cells are determined. Finally, we use PHAST sweeps (restricted to active cells) to compute distances to vertices of active cells. We now describe the (generic) preprocessing of isoPHAST, before we present strategies to determine active cells.

First, we compute a (single-level) partition $\mathcal{V} = \{V_1, \ldots, V_k\}$ of the graph. Then, we use CH to contract all cell-induced subgraphs, but leave boundary vertices of the partition *uncontracted*. Afterwards, we reorder vertices in the graph such that uncontracted *core* vertices are pushed to the front, breaking ties by cell (providing the same benefits as in CRP). We order contracted vertices within cells by their CH levels (as required by PHAST, see Section 2). As a result of preprocessing, we obtain a (single) *upward graph* $G^\uparrow$, containing (for all cells) edges to higher-ranked vertices, added shortcuts between core vertices, and all boundary edges. Conversely, the *downward graph* $G^\downarrow$ stores only downward edges (i.e., boundary vertices have no incoming edges). Further steps of preprocessing depend on the query strategy and are described below.

**isoPHAST-CD.** Our first strategy (CoreDijkstra) executes isoDijkstra on the core graph to determine active cells. This requires eccentricities for core vertices, which are efficiently obtained from preprocessed data. To compute $\text{ecc}(u)$ of some vertex $u$, we run (as a last step of preprocessing) Dijkstra's algorithm on the core graph (restricted to the cell $V_i$ containing $u$), followed by a PHAST sweep over the internal vertices of $V_i$. Whenever processing a vertex $v$ (of $V_i$), we update the eccentricity by setting $\text{ecc}(u) = \max\{\text{ecc}(u), d(v)\}$.

A query starts by running (iso)Dijkstra from the source $s$ in $G^\uparrow$. Within the cell of $s$, this corresponds to an upward CH search, since $G^\uparrow$ stores only upward edges. At core vertices, we update flags $\mathtt{i}(\cdot)$ and $\mathtt{o}(\cdot)$ to determine active cells (as described in Section 4, using an adjacency array to store vertices required for additional checks, similar to isoGRASP). In the special case that the core is not reached, only the cell of $s$ is set active. Then, we process every active cell by running PHAST on its internal vertices. Thereby, we obtain distances to all vertices that are both in range and contained in an active cell (due to the stopping criterion, we get upper bounds for vertices out of range, which suffices for isochrone detection).

Regarding output, boundary (isochrone) edges are found by isoDijkstra. Isochrone edges connecting internal vertices are obtained in linear sweeps. When settling a vertex $v$, higher-ranked neighbors have final distance labels. The label of $v$ is final after scanning incoming edges $(u, v) \in G^\downarrow$. Thus, looping through incoming edges a second time suffices to find isochrone edges after a few modifications to $G^\downarrow$. First, as only original edges are relevant candidates, we indicate this with an additional flag (per edge). Second, (directed) edges $(v, u) \in E$ to vertices $u$ of higher rank are (in general) not contained in $G^\downarrow$. To ensure that adjacent vertices (wrt. $G$) are also adjacent in $G^\downarrow$, we add dummy edges (with a length of infinity) to $G^\downarrow$.

**isoPHAST-CP.** Instead of isoDijktra, this strategy (CorePHAST) runs PHAST within the core. We compute eccentricities after the (generic) preprocessing (as described above). Next, CH preprocessing is run on the core, too. We reorder core vertices according to CH levels. Finally, we update $G^\uparrow$ and $G^\downarrow$ (obtained after cell contraction), adding core shortcuts and removing the obsolete boundary edges of the partition (that were needed to compute eccentricites).

Queries strictly follow the three-phase pattern discussed above. First, we perform a forward CH search in $G^\uparrow$, until the priority queue is empty. Second, we determine active cells and compute correct distance labels for all core vertices. To achieve this, we run a PHAST sweep over the core vertices (that were sorted accordingly), relaxing edges in $G^\downarrow$ to propagate distance values to vertices of lower rank. Again, we maintain flags, $\texttt{i}(\cdot)$ and $\texttt{o}(\cdot)$, and an adjacency array for vertex checks to determine active cells. To find isochrone edges between boundary vertices on-the-fly, we add dummy edges and edge flags inside the core (similar to the third phase of isoPHAST-CD). The third phase (sweeps over active cells) is identical to isoPHAST-CD.

**isoPHAST-DT.** Our last strategy (DistanceTable) uses a *distance (bounds) table* to speed up the second phase (determining active cells). Working with such tables (instead of a dedicated core search) benefits from *edge partitions*, since the unique assignment of edges to cells simplifies isochrone edge retrieval (we avoid checking for boundary edges in the second phase). Given a partition $\mathcal{E} = \{E_1, \ldots, E_k\}$ of the edges of $G$, the table stores for each pair $E_i, E_j$ of cells a lower bound $\underline{d}(E_i, E_j)$ and an upper bound $\overline{d}(E_i, E_j)$ on the distance from $E_i$ to $E_j$, i.e., $\underline{d}(E_i, E_j) \leq d(u, v) \leq \overline{d}(E_i, E_j)$ for all $u \in E_i$, $v \in E_j$ (we abuse notation, saying $v \in E_i$ if $v$ is an endpoint of at least one edge $e \in E_i$). Given a source $s \in E_i$ (if $s$ is ambiguous, pick any cell containing $s$) and a limit $\tau$, cells $E_j$ with $\underline{d}(E_i, E_j) \leq \tau < \overline{d}(E_i, E_j)$ are set active.

To (quickly) compute (not necessarily tight) distance bounds, we make use of the data computed during generic preprocessing (that is now based on an edge partition). On the resulting core graph we run, for each cell $E_i$, a (multi-source) forward CH search from all boundary vertices in $E_i$. Next, we perform a PHAST sweep on $G^\downarrow$ (i.e., the *complete* downward graph), keeping track of the minimum and maximum distance label per target cell. This yields, for all cells, lower bounds $\underline{d}(E_i, \cdot)$, and upper bounds on the distance from *boundary* vertices of $E_i$ to each cell. To obtain the desired bounds $\overline{d}(V_i, \cdot)$, we increase these values by the *(backward) boundary diameter* of $E_i$, i.e., the maximum distance from any vertex in $E_i$ to a boundary vertex in $E_i$. To compute the diameter of a cell $E_i$, we run, for each boundary vertex $v \in E_i$, Dijkstra's algorithm on the *reverse* core graph (i.e., inverting edge directions), until all boundary vertices of $E_i$ were scanned (this can be easily verified using a counter). Using a PHAST sweep over the downward graph of $E_i$, we obtain the unlimited (i.e., not limited to $E_i$) backward eccentricity of $v$. The diameter of $V_i$ is the maximum eccentricity of its boundary vertices. Next, we perform the actual CH preprocessing on $G$, using a purely greedy order (boundary vertices are *not* artificially kept on top of the contraction order). This results in a better contraction order and sparser graphs $G^\uparrow$ and $G^\downarrow$, reducing preprocessing and query times. Finally, we extract (and store) the relevant search graph $G_i^\downarrow$ for each $E_i \in \mathcal{E}$. This requires an RPHAST selection phase per cell, using all (ambiguous and distinct) vertices of a cell as input.

Queries start with a forward CH search on $G^\uparrow$ (applying no stopping criterion). Next, active cells are determined in a linear sweep over one row (corresponding to the source cell) of the distance table. The third phase runs an RPHAST sweep for each active cell, obtaining distances to all its vertices [7]. Although vertices can be contained in several search graphs, we do not reinitialize distance labels between sweeps (the source remains unchanged). To obtain isochrone edges, we proceed as before, looping through incoming downward edges twice (again, we add dummy edges to $G_i^\downarrow$ for correctness). To avoid duplicates in the output (due to vertices contained in several search graphs), edges in $G_i^\downarrow$ get an additional flag to indicate whether the edge belongs to the corresponding cell $E_i$. Isochrone edges are added to the output only if this flag is set.

As mentioned above, search graphs may share some vertices (e.g., the vertex with maximum CH rank is contained in every search graph). Vertices contained in many graphs increase memory consumption and slow down queries. Therefore, we use *search graph compression*, i.e., we store the topmost vertices of the hierarchy (and their incoming edges) in a separate graph $G_c^\downarrow$ and remove them from all search graphs $G_i^\downarrow$. During queries, we first perform a linear sweep over $G_c^\downarrow$ (obtaining distances for all $v \in G_c^\downarrow$), before processing search graphs of active cells.

**Parallelization.** The first steps of preprocessing are executed in parallel, namely, building cell graphs, contracting internal vertices, inserting dummy edges, and reordering internal vertices by level. Afterwards, threads are synchronized. The upward and downward graphs are computed sequentially. Finally, computation of eccentricities (or cell diameters) is parallelized again. Since threads operate on distinct vertex sets, there are no concurrent accesses to distance labels, and we can share labels among threads. However, each thread needs its own priority queue. All remaining steps of preprocessing (depending on the strategy) are executed sequentially.

As for queries, the first two phases are run sequentially. Both the CH search and isoDijkstra are difficult to parallelize. Running PHAST (on the core) in parallel does not pay off (the core is rather dense, resulting in many CH levels). Distance table operations, on the other hand, are very fast, so parallelization is not necessary. In the third phase, however, active cells can be assigned to different threads. Again, we share distance labels among threads, except for isoPHAST-DT (search spaces may overlap, so each thread uses own labels to avoid concurrent accesses). As a result, in isoPHAST-DT each thread runs a forward CH search to initialize its labels. We store a copy of the graph $G^\downarrow$ once per NUMA node for faster access during queries.

**Table 1.** Performance of basic one-to-all and one-to-many building blocks. Execution times are sequential. One-to-many algorithms compute distances to all targets in a ball of size $|B| = 2^{14}$.

| Algorithm | [our] | [7, 15] |
|---|---:|---:|
| Dijkstra (1-to-all) | 2 653.18 | – |
| PHAST | 144.16 | 136.92 |
| GRASP | 171.11 | 169.00 |
| Dijkstra (1-to-many) | 7.34 | 7.43 |
| RPHAST (selection) | 1.29 | 1.80 |
| RPHAST (query) | 0.16 | 0.17 |

Running the third phase in parallel can make the second phase of isoPHAST-CP a bottleneck. Therefore, we alter the way of computing $\mathtt{i}(\cdot)$ and $\mathtt{o}(\cdot)$. When settling a vertex $v \in V_i$, we set $\mathtt{o}(V_i)$ if $d(v) + \mathrm{ecc}(v) > \tau$, and $\mathtt{i}(V_i)$ if $d(v) \leq \tau$. These checks are less accurate (marking more cells), but we no longer have to check unreachable boundary vertices (wrt. $V_i$). Observe that correctness of isoPHAST-CP is maintained (no stopping criterion is applied and $\max_{v \in V_i}(d(v) + \mathrm{ecc}(v))$ is a valid upper bound on the distance to each vertex in $V_i$, so no active cells are missed).

## 6. Experiments

Our code is written in C++ (using OpenMP for parallelization), compiled with g++ 4.8 (-O3). The input for all experiments is the road network of Western Europe made available for the 9th DIMACS Implementation Challenge ($|V| = 18\,010\,173, |E| = 42\,188\,664$), with travel time (in seconds) as edge lengths [10]. For isoCRP and isoGRASP, we use a 4-level partition obtained from PUNCH [6], with maximum cell sizes of $2^8, 2^{12}, 2^{16}, 2^{20}$, respectively. For (single-level) vertex partitions required by isoPHAST, we used BUFFOON [27]. Edge partitions for isoPHAST-DT were retrieved following the approach described in [26, 28]. Our CH implementation follows [19], but adopting tuning parameters (e. g. for priority terms and staged hop limits) from [4]. Our GRASP implementation includes implicit initialization [14] and (downward) edge reduction [16]. As opposed to [14], we resort to a 4-level partition, with only minor effects on running times in preliminary experiments (similar observations were made in [15]). For comparability, Table 1 shows running times of our implementations of the basic batched query algorithms on a single core of a 4-core Intel Xeon E5-1630v3 clocked at 3.7 GHz, with 128 GiB of DDR4-2133 RAM, 10 MiB of L3 and 256 KiB of L2 cache (chosen because it most closely resembles the machines used in [7, 15]). We report average running time over 1 000 random one-to-all queries for Dijkstra, PHAST, and GRASP. For one-to-many techniques (Dijkstra and RPHAST), we follow a procedure of [7], picking a center vertex $c$ at random and running Dijkstra's algorithm from $c$ until $|T|$ vertices were scanned, making them the target set $T$. Distances from a random source $s \in T$ to all vertices in $T$ are computed. For RPHAST, we report both target selection (sel) and query (qry) time. We also report numbers (as-is) from [7, 15] for comparison. Even when accounting for hardware differences, we see that query times of our implementations are similar to original publications (target selection of RPHAST is even slightly faster).

**Table 2.** Performance of sequential queries. We report customization time (excluding metric-independent partitioning) and space requirements (space per additional metric given in brackets, if it differs). The table shows the average number of settled vertices (#Settled, in thousands) and running times of queries, for time limits $\tau = 100$ and $\tau = 500$. Best values (except Dijkstra wrt. space) are highlighted in bold.

| Algorithm | Custom Time [s] | Custom Space [MiB] | Queries ($\tau = 100\,\text{min}$) #Settled | Queries ($\tau = 100\,\text{min}$) Time [ms] | Queries ($\tau = 500\,\text{min}$) #Settled | Queries ($\tau = 500\,\text{min}$) Time [ms] |
|---|---|---|---|---|---|---|
| RangeDijkstra | – | 646 | 460 k | 68.32 | 7 041 k | 1 184.06 |
| isoCRP | **1.70** | 900 (138) | 101 k | 15.44 | 354 k | 60.67 |
| isoGRASP | 2.50 | 1 856 (1 094) | 120 k | 10.06 | 387 k | 37.77 |
| isoPHAST-CD | 26.11 | 785 | 440 k | **6.09** | 1 501 k | 31.63 |
| isoPHAST-CP | 1 221.84 | **781** | 626 k | 15.02 | 2 029 k | 31.00 |
| isoPHAST-DT | 1 079.11 | 2 935 | 597 k | 9.96 | 1 793 k | **24.80** |

Regarding isoPHAST, we performed preliminary studies to obtain reasonable parameters for partition sizes ($k$) and search graph compression ($|G_c^{\downarrow}|$, denoting the number of vertices in the search graph for isoPHAST-DT). For sequential CD (CP) queries, a value of $k = 2^{12}$ ($2^{11}$) yields best query times. For fewer cells (i.e., coarser partitions), the third phase scans a large portion of the graph and becomes the bottleneck. Using a more fine-grained partition yields a larger core graph, slowing down the second phase. Consequently, a smaller value for $k$ (256) becomes favorable when queries are executed in parallel (as the third phase becomes faster). For DT, similar effects occur for different values of $k$. Additionally, subgraph compression has a major effect on running times (and space consumption). For small sizes of $G_c^{\downarrow}$, high-ranked vertices occur in multiple cells, but a larger graph $G_c^{\downarrow}$ leads to unnecessary vertex scans during queries. Setting $k = 2^{14}(2^{12})$ and $|G_c^{\downarrow}| = 2^{16}(2^{13})$ resulted in fastest sequential (parallel) queries for isoPHAST-DT.

Table 2 shows performance of all algorithms discussed in Sections 4 and 5, reporting figures on customization and sequential queries. Experiments were conducted on a dual 8-core Intel Xeon E5-2670 clocked at 2.6 GHz, with 64 GiB of DDR3-1600 RAM, 20 MiB of L3 and 256 KiB of L2 cache. Customization is executed in parallel (using all 16 cores). Timings for PHAST include all steps of preprocessing except computing the partition (since it is independent of the metric). For queries, we report average running times of 1 000 random queries for mid-range ($\tau = 100$) and long-range time limits ($\tau = 500$, this was the hardest limit for most approaches, as it maximized the number of active cells). As expected, techniques based on MLD provide better customization times, while isoPHAST achieves the lowest query times (CD for mid-range and DT for long-range queries, respectively). Customization of isoCRP and isoGRASP is very practical (below three seconds). For isoPHAST-CD, the lightweight preprocessing pays of as well, allowing customization in less than 30 seconds. The comparatively expensive preprocessing times of isoPHAST-CP and DT are dominated by expensive core contraction. Still, (metric-dependent) preprocessing is far below half an hour, which is suitable for applications that do not require realtime metric updates. Compared to isoCRP, isoGRASP requires almost an order of magnitude of additional space for the downward graph (having about 110 million edges). Even when executed sequentially, all approaches take well below 100 ms, which is significantly faster than isoDijkstra. The number of settled vertices is significantly larger for isoPHAST (but data access is more cache friendly). Query times of isoPHAST are below MLD techniques for both limits, with the exception of isoPHAST-CP for small ranges (because the whole core graph is scanned). Again, the performance of isoPHAST-CD is quite noteable, providing the fastest queries for (reasonable) mid-range limits and decent query times for the long-range limit. Finally, query times of isoPHAST-DT show best scaling behavior, with lowest times for harder queries.

**Table 3.** Performance of parallel queries. We report the same figures as in Table 2, but queries use 16 threads. Best values (per column) are highlighted in bold.

| Algorithm | Custom Time [s] | Space [MiB] | Queries ($\tau = 100\,\text{min}$) # Settled | Time [ms] | Queries ($\tau = 500\,\text{min}$) # Settled | time [ms] |
|---|---|---|---|---|---|---|
| isoCRP | **1.70** | 900 (138) | 100 k | 2.73 | 354 k | 7.86 |
| isoGRASP | 2.50 | 1 856 (1 094) | 120 k | 2.35 | 387 k | 5.93 |
| isoPHAST-CD | 38.07 | 769 | 918 k | **1.61** | 4 578 k | 8.22 |
| isoPHAST-CP | 1 432.39 | **766** | 944 k | 4.47 | 5 460 k | 7.86 |
| isoPHAST-DT | 865.50 | 1 066 | 914 k | 1.74 | 2 979 k | **3.80** |

Table 3 shows parallel times for the same set of random queries, using the same machine. Note that preprocessing time of isoPHAST changes due to different parameter choices. Most approaches scale very well with the number of threads, providing a speedup of (roughly) 8 using 16 threads. Note that factors (according to tables) are much lower for isoPHAST, since we use tailored partitions for single-thread queries. In fact, isoPHAST-DT scales best when run on the same preprocessed data (speedup of 11), since its sequential workflow (upward CH search, table lookups) is very fast. Considering MLD techniques, isoGRASP scales worse than isoCRP (speedup of 6.5 compared to 7.7), probably because it is memory bandwidth bounded (while isoCRP comes with more computational overhead). Consequently, isoGRASP benefits greatly from storing a copy of the downward graph on each NUMA node. As one may expect, speedups are slightly lower for mid-range queries ($\tau = 100$). The isoPHAST approaches yield best query times, below 2 ms for mid-range queries, and below 4 ms for the long-range limit. In summary, all algorithms enable queries that are fast enough for practical applications, achieving speedups by more than two orders of magnitude compared to isoDijkstra.

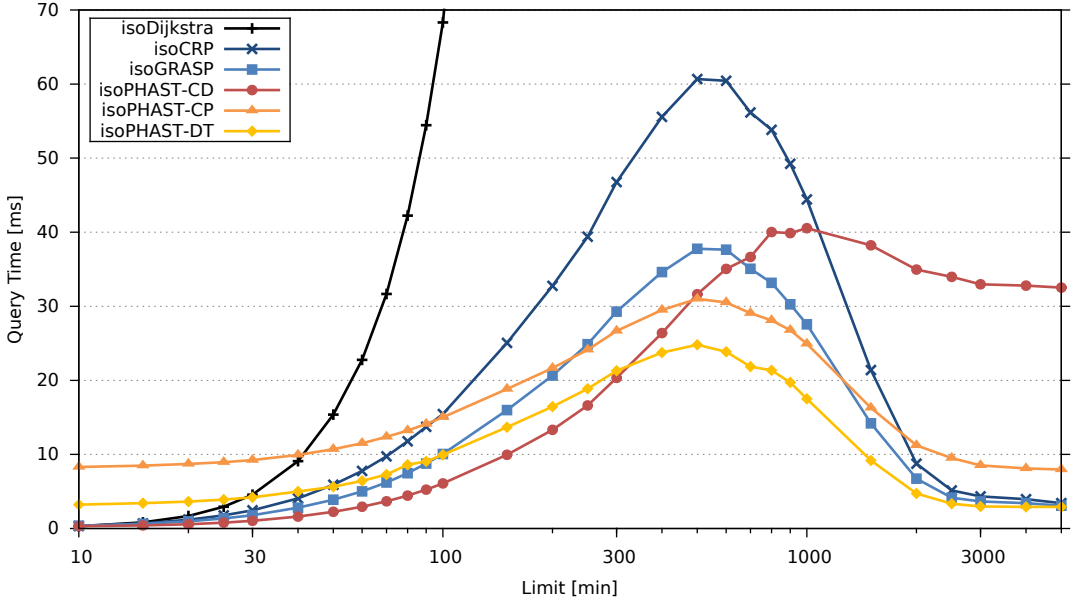

**Figure 1.** Sequential query times for varying time limits, ranging from 10 to (roughly) 4 700 minutes (the diameter of our input graph). Each point in the plot is the average of 1 000 random queries.

Figure 1 shows how (sequential) query times scale for varying time limits. Timings of all algorithms (except isoDijkstra) follow a characteristic curve. Query times first increase with the limit $\tau$ (the isoline is extended, cutting more active cells), before dropping again once $\tau$ exceeds 500 (the isochrone reaches the boundary of the network, decreasing the number of active cells). For values $\tau > 4\,710$ (minutes), all vertices are in range, making queries very fast (as there are no active cells). For small values of $\tau$, the MLD techniques and isoPHAST-CD are fastest. Conversely, isoPHAST-CP is slowed down by the linear sweep over the core graph (taking about 6 ms, independent of the value $\tau$), while isoPHAST-DT suffers from distance bounds in the table that are not tight. On the other hand, the second phase of isoPHAST-CD has no stopping criterion, making it the slowest variant for large values of $\tau$ (while the other isoPHAST variants benefit from good scaling behavior). Regarding MLD techniques, isoGRASP is up to almost twice as fast as isoCRP, providing a decent trade-off between customization effort and query times. In the parallel query setting (not reported in the figure), query times follow the same characteristic curves. The linear sweep in the second phase of isoPHAST-CP becomes slightly faster, since the core graph is smaller (due to a different partition). Also ,the performance gap between isoCRP and isoGRASP is slightly smaller when using multiple threads.

## 7. Final Remarks

We proposed a formal definition of the isochrone problem, asking for a compact representation of the region in a road network that is in range (given some resource limit). We defined output that is reasonable, e. g., for actual *visualization* of isochrones. The approach of [25] (used as state-of-the-art in recent work [13]) can be modified easily to take isochrone edges as input. Given a (planar) graph representation of the network, this can be achieved by traversing the faces of the graph along the border of the region in range, switching to the corresponding twin face at isochrone edges. Preliminary experiments showed that extracting the isochrone region (according to the definition in [25]) this way takes only tens of milliseconds. While the desired output may differ for other applications, our algorithms can be readily extended to most reasonable formats.

We introduced a portfolio of speedup techniques for the isochrone problem. While no single approach is best in all criteria (preprocessing effort, space, query time, simplicity), the right choice depends on the application and its requirements. If user-dependent metrics are needed, the fast and lightweight customization isoCRP is helpful. Fast query times subject to frequent updates (e. g., due to live traffic) are enabled by isoGRASP. If customization time below a minute is acceptable and ranges are low, isoPHAST-CD is a candidate that may achieve even faster query times. The remaining variants of isoPHAST show best scaling behavior, making them suitable for long-range isochrones on large graphs, or if customizability is not required.

Current and future work includes algorithms for computing actual isolines that are fast enough for interactive applications (exploiting the techniques presented here). Moreover, we are interested in visualizing the cruising range of electric vehicles, taking both travel time and energy consumption into account. Finally, we investigate the possibility of customizable isoPHAST (exploiting, e. g., CCH [11]), and variants that do not (explicitly) require a partition of the graph.

# A. Omitted Figures

Figure 2 shows two examples for which a naïve MLD approach may produce wrong outputs (see also the challenges described in Section 4). Below, Figure 3 illustrates the work of the isoPHAST algorithms described in Section 5.

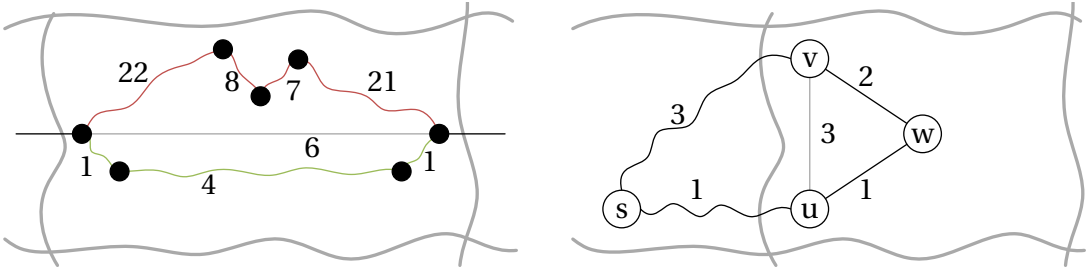

**Figure 2.** Two examples for which a naïve approach produces wrong outputs. Gray edges denote shortcut edges. Left: Imagine a mountain in the middle of the cell. The shortcut edge from the left to the right boundary vertex corresponds to a fast road through a tunnel (the green edges) and can be traversed within the time limit. However, this does not imply that we can reach the summit of the mountain via a mountain road (the red edges). Thus, without using eccentricites, it may happen that we do not descend into the cell and hence miss an isochrone edge on the mountain road. Right: Assume a limit $\tau = 4$. When we scan $u$, we can traverse the shortcut edge $(u, v)$, since the distance at $u$ plus the shortcut length equals the limit. Thus, we do not descend into the cell at $u$, but we relax the outgoing shortcut edge $(u, v)$. Next, scan $v$, where the sum of distance at $v$ and the shortcut length is greater than the $\tau$. Therefore, we descend into the cell and relax the edge $(v, w)$. Since $w$ has a distance label greater than $\tau$ and the queue contains no other vertices, the search stops. Now, we erroneously report $(v, w)$ as an isochrone edge, since the distance label at $v$ is below $\tau$, while the distance label of $w$ is greater than $\tau$. The $s-w$ path via $u$ that makes $w$ reachable within $\tau$ is not found. This motivates out two-phase approach (which ensures that distance labels in cells are correct).

# B. Omitted Experiments

This section contains experiments omitted in the main document. Below, we present alternative variants of isoCRP and isoGRASP, using different strategies to determine active cells. We also include detailed experiments considering parameter tuning of isoPHAST variants.

**Alternative Multilevel Strategies.** Given a boundary vertex $u$ on some cell $V_i^\ell$ at level $\ell$, we define its *unrestricted (level-$\ell$) eccentricity*, denoted $\text{ecc}'_\ell(u)$, as the maximum distance $d(u, v)$ to a vertex $v \in V_i^\ell$ (in $G$) in the same cell (i.e., distances are not restricted to cell-induced subgraphs). We compute, for every boundary vertex $v$, and for each level $\ell$, the unrestricted level-$\ell$ eccentricity of $v$. The procedure described in Section 4 yields upper bounds on (unrestricted) eccentricities if the corresponding subgraphs are strongly connected (otherwise, we just set the eccentricity to $\infty$).

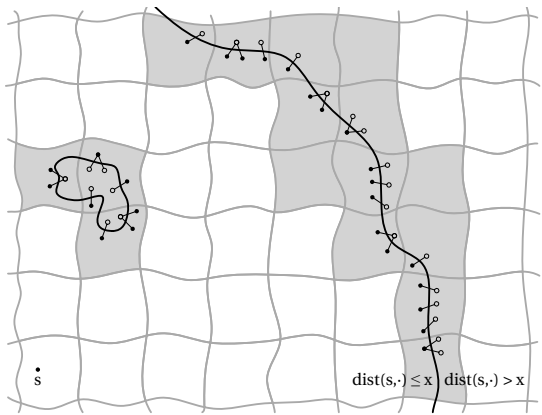

**Figure 3.** Illustration of the isoPHAST algorithms. Thick black lines denote the isoline, which is intersected by isochrone edges (thin black lines). Cells that are colored gray denote active cells. Note that there are cells that are marked active despite containing no isochrone edges (however, this does not violate correctness). Also note that besides the active cells along the primary isoline, there may be active cells in the interior of the area bounded by the primary isoline, for example due to mountains whose summits we cannot reach within the time limit.

During queries, we may now check for boundary vertices whether the sum of their distance label and their eccentricity exceeds the time limit. If it does, we mark it as an active cell. We refer to this approach (which is simpler but less accurate than the approach of Section 4) as *isoCRP (none)*. Since we compute upper bounds on eccentricities during customization, this approach may falsely mark cells as active. While this does not violate correctness, it can significantly slow down the second phase of the query. This holds in particular for cells that induce subgraphs which are not strongly connected. Recall that eccentricities are computed on this subgraph, and hence we assign eccentricities of $\infty$ to some vertices. In other words, the corresponding cell will always be marked active if at least one of its boundary vertices is in range. Also note that a significant number of cells induces subgraphs that are not strongly connected (think of cuts at directed edges representing lanes of a highway). Therefore, we present an approach to improve upper bounds on eccentricities. We run customization as described above, but add a second sweep over all boundary vertices. For each boundary vertex, we run another Dijkstra search, but no longer limit it to the current cell (note that outside of cells, we can make use of all preprocessed shortcuts, even on higher levels). This search is aborted once that all vertices of the cell were reached (we can use a counter to easily check this condition). This yields our second approach, *isoCRP*

**Table 4.** Performance of (sequential) MLD query variants.

| | CUSTOM | | | QUERIES | | | |
| | | | limit $\tau = 100\,\text{min}$ | | limit $\tau = 500\,\text{min}$ | |
| algorithm | time [s] | space [MiB] | # scans | time [ms] | # scans | time [ms] |
| --- | --- | --- | --- | --- | --- | --- |
| isoCRP (none) | 1.70 | 900 (138) | 200 020 | 26.12 | 1 956 102 | 267.40 |
| isoCRP (all) | 8.42 | 900 (138) | 122 550 | 17.33 | 436 301 | 69.62 |
| isoCRP (inf) | 4.91 | 900 (138) | 133 276 | 18.48 | 492 876 | 77.46 |
| isoCRP (scc) | 2.48 | 900 (138) | 140 344 | 19.34 | 544 784 | 85.17 |
| isoCRP (updown) | 1.70 | 900 (138) | 100 789 | 15.44 | 354 291 | 60.67 |
| isoGRASP (none) | 2.50 | 1 856 (1 094) | 215 302 | 13.99 | 1 846 034 | 123.01 |
| isoGRASP (all) | 9.65 | 1 856 (1 094) | 150 079 | 10.59 | 499 641 | 40.88 |
| isoGRASP (inf) | 5.85 | 1 856 (1 094) | 157 614 | 10.84 | 543 677 | 43.31 |
| isoGRASP (scc) | 3.26 | 1 856 (1 094) | 163 483 | 11.42 | 588 081 | 48.06 |
| isoGRASP (up) | 2.50 | 1 856 (1 094) | 156 766 | 11.38 | 547 253 | 44.62 |
| isoGRASP (updown) | 2.50 | 1 856 (1 094) | 120 327 | 11.56 | 387 053 | 45.42 |
| isoGRASP (sep) | 2.69 | 1 879 (1 117) | 120 327 | 10.06 | 387 053 | 37.77 |

*(all)*. To speed up customization (at the cost of slightly increased query times), we only run this search for boundary vertices that have an infinite eccentricity after the first run (*isoCRP (inf)*). In fact, we can even try to make use of eccentricities previously updated during the second run: before running a Dijkstra search for a vertex $u$ with $\text{ecc}_\ell(u) = \infty$, we check all outgoing shortcuts $(u, v)$ (on the current level $\ell$). If $\text{len}((u,v)) + \text{ecc}_\ell(v) \neq \infty$, we can use this sum as finite upper bound and omit the Dikstra search, yielding *isoCRP (scc)*.

Table 4 shows (sequential) query timings of different MLD variants (for 1 000 random queries). The first block corresponds to isoCRP queries based on unrestricted eccentricites. Next, isoCRP (updown) corresponds to the approach presented in the main document. For isoGRASP, similar figures are reported. We also distinguish several variants of the approach described in the main document. In particular, (up) uses unrestricted eccentricites only in the second phase, (updown) uses restricted eccentricites (as described in Section 4), but does not store vertices for additional checks in a dedicated adjacency array (instead, we scan matrices of the overlay). Finally, (sep) is the variant described in the main text. Table 5 shows figures for parallel queries.

Table 5. Performance of (parallel) MLD query variants.

|  | CUSTOM | | QUERIES | | | |
|  | | | limit $x = 100\,\text{min}$ | | limit $x = 500\,\text{min}$ | |
| algorithm | time [s] | space [MiB] | # scans | time [ms] | # scans | time [ms] |
| --- | --- | --- | --- | --- | --- | --- |
| isoCRP (none) | 1.70 | 900 (138) | 200 020 | 3.92 | 1 956 102 | 25.02 |
| isoCRP (all) | 8.42 | 900 (138) | 122 550 | 2.84 | 436 301 | 9.08 |
| isoCRP (inf) | 4.91 | 900 (138) | 133 276 | 2.99 | 492 876 | 8.87 |
| isoCRP (scc) | 2.48 | 900 (138) | 140 344 | 3.02 | 544 784 | 9.36 |
| isoCRP (updown) | 1.70 | 900 (138) | 100 789 | 2.73 | 354 291 | 7.86 |
| isoGRASP (none) | 2.50 | 1 856 (1 094) | 215 302 | 2.73 | 1 846 034 | 13.59 |
| isoGRASP (all) | 9.65 | 1 856 (1 094) | 150 079 | 2.35 | 499 641 | 6.04 |
| isoGRASP (inf) | 5.85 | 1 856 (1 094) | 157 614 | 2.40 | 543 677 | 6.37 |
| isoGRASP (scc) | 3.26 | 1 856 (1 094) | 163 483 | 2.47 | 588 081 | 6.97 |
| isoGRASP (up) | 2.50 | 1 856 (1 094) | 156 766 | 2.37 | 547 253 | 6.39 |
| isoGRASP (updown) | 2.50 | 1 856 (1 094) | 120 327 | 2.50 | 387 053 | 6.19 |
| isoGRASP (sep) | 2.69 | 1 879 (1 117) | 120 327 | 2.35 | 387 053 | 5.93 |

**Table 6.** Performance of isoPHAST-CD and isoPHAST-CP for varying number of cells. Queries are executed sequentially (seq.) and in parallel (par.) and given in milliseconds. The variant with the least absolute (relative) deviations is highlighted in dark (light) gray.

| algorithm | exec. | $\tau$ | # cells | | | | | | | |
|---|---|---|---|---|---|---|---|---|---|---|
| | | | 128 | 256 | 512 | 1 024 | 2 048 | 4 096 | 8 192 | 16 384 |
| isoPHAST-CD | seq. | 10 | 1.92 | 1.17 | 0.74 | 0.51 | 0.38 | 0.32 | 0.33 | 0.41 |
| | | 100 | 12.51 | 10.37 | 8.88 | 7.75 | 6.83 | 6.09 | 5.63 | **5.56** |
| | | 500 | 65.45 | 51.81 | 42.24 | 35.14 | 31.87 | 31.24 | 33.70 | 40.03 |
| | par. | 10 | 1.87 | 0.91 | 0.59 | 0.34 | 0.40 | 0.43 | 1.13 | 2.08 |
| | | 100 | 2.26 | 1.62 | 1.34 | 1.33 | 1.50 | 1.83 | 2.63 | 4.10 |
| | | 500 | 8.30 | 8.19 | 9.02 | 10.51 | 13.18 | 17.27 | 23.58 | 38.68 |
| isoPHAST-CP | seq. | 10 | **4.36** | 4.50 | 5.08 | 6.35 | 8.25 | 10.86 | 14.56 | 19.30 |
| | | 100 | 15.30 | 14.05 | **13.69** | 13.98 | 14.93 | 16.59 | 19.19 | 23.21 |
| | | 500 | 68.26 | 54.21 | 43.40 | 35.41 | 31.10 | 28.44 | **28.41** | 30.64 |
| | par. | 10 | 3.85 | 3.68 | 4.68 | 4.69 | 5.87 | 7.94 | 10.41 | 13.77 |
| | | 100 | **4.33** | 4.44 | 4.80 | 5.56 | 6.77 | 8.60 | 11.10 | 14.16 |
| | | 500 | 8.53 | 7.90 | **7.74** | 7.97 | 8.83 | 10.14 | 12.10 | 15.27 |

**Parameter Tuning for isoPHAST.** The following Table 6 shows running times of isoPHAST-CD and isoPHAST-CP for partitions with different numbers of cells. We report sequential and parallel running times for different time limits (given in minutes). Table 7 shows sequential query times obtained for different choices of partition and compressed subgraph size. Table 8 shows parallel running times for the same parameter set.

**Table 7.** Sequential query times [ms] of isoPHAST-DT for varying numbers of cells in underlying partitions and compressed subgraph sizes $|G_\mathrm{c}^\downarrow|$. Query times are reported for different time limits (given in minutes) The variant with the least absolute (relative) deviations is highlighted in dark (light) gray.

| | | # CELLS | | | | | | | |
|---|---|---|---|---|---|---|---|---|---|
| $|G_\mathrm{c}^\downarrow|$ | $\tau$ | 128 | 256 | 512 | 1 024 | 2 048 | 4 096 | 8 192 | 16 384 |
| 128 | 10 | 11.23 | 6.50 | 4.14 | 2.88 | 2.40 | 2.42 | 2.80 | 3.32 |
| | 100 | 28.96 | 23.51 | 21.98 | 22.39 | 26.56 | 37.02 | 56.37 | 85.55 |
| | 500 | 111.79 | 98.99 | 95.56 | 94.05 | 105.11 | 134.85 | 195.08 | 280.13 |
| 256 | 10 | 11.13 | 6.39 | 4.04 | 2.77 | 2.29 | 2.23 | 2.49 | 3.07 |
| | 100 | 28.87 | 23.23 | 21.17 | 21.41 | 25.19 | 33.99 | 50.37 | 77.16 |
| | 500 | 109.61 | 96.93 | 91.62 | 89.42 | 99.34 | 122.52 | 174.39 | 250.95 |
| 512 | 10 | 10.97 | 6.50 | 3.86 | 2.60 | 2.09 | 1.95 | 2.13 | 2.63 |
| | 100 | 28.11 | 22.36 | 19.93 | 19.55 | 22.26 | 28.78 | 41.79 | 63.52 |
| | 500 | 108.24 | 92.88 | 85.50 | 79.81 | 84.85 | 100.98 | 138.64 | 200.31 |
| 1 024 | 10 | 10.77 | 5.99 | 3.66 | 2.38 | 1.88 | 1.65 | 1.68 | 2.02 |
| | 100 | 27.44 | 21.16 | 18.45 | 17.16 | 18.83 | 22.84 | 29.75 | 44.99 |
| | 500 | 104.46 | 87.64 | 77.81 | 68.71 | 70.07 | 75.56 | 94.72 | 132.81 |
| 2 048 | 10 | 10.62 | 5.85 | 3.52 | 2.27 | 1.73 | 1.47 | 1.41 | 1.54 |
| | 100 | 26.96 | 20.31 | 17.11 | 15.42 | 15.61 | 17.41 | 20.76 | 27.35 |
| | 500 | 102.93 | 83.64 | 71.59 | 61.26 | 57.49 | 57.40 | 64.10 | 78.22 |
| 4 096 | 10 | 10.58 | 5.84 | 3.49 | 2.27 | 1.71 | 1.43 | **1.34** | 1.39 |
| | 100 | 26.73 | 19.77 | 16.23 | 14.49 | 14.07 | 14.15 | 15.32 | 18.17 |
| | 500 | 100.66 | 81.07 | 67.56 | 55.98 | 49.85 | 46.75 | 46.39 | 51.84 |
| 8 192 | 10 | 10.72 | 5.96 | 3.61 | 2.40 | 1.82 | 1.53 | 1.42 | 1.43 |
| | 100 | 26.46 | 19.50 | 15.75 | 13.42 | 12.31 | 11.83 | 11.79 | 12.68 |
| | 500 | 99.99 | 79.52 | 65.14 | 52.90 | 44.87 | 39.33 | 36.61 | 36.31 |
| 16 384 | 10 | 11.52 | 6.25 | 3.91 | 2.69 | 2.12 | 1.83 | 1.75 | 1.81 |
| | 100 | 26.33 | 19.56 | 15.70 | 13.37 | 11.87 | 10.92 | 10.32 | 10.51 |
| | 500 | 101.44 | 79.08 | 63.83 | 50.95 | 42.51 | 35.80 | 32.17 | 29.02 |
| 32 768 | 10 | 11.96 | 6.84 | 4.46 | 3.26 | 2.69 | 2.39 | 2.28 | 2.62 |
| | 100 | 26.94 | 19.94 | 16.05 | 13.50 | 12.03 | 11.17 | 10.09 | **9.72** |
| | 500 | 97.85 | 78.36 | 63.57 | 50.28 | 41.49 | 34.30 | 29.35 | 26.66 |
| 65 536 | 10 | 12.48 | 7.81 | 5.44 | 4.22 | 3.65 | 3.36 | 3.26 | 3.24 |
| | 100 | 27.54 | 21.01 | 17.47 | 14.34 | 12.76 | 11.85 | 10.64 | 10.03 |
| | 500 | 98.19 | 79.07 | 63.54 | 50.44 | 41.63 | 34.07 | 28.80 | 25.03 |

**Table 8.** Parallel query times [ms] of isoPHAST-DT for varying numbers of cells in underlying partitions and compressed subgraph sizes $|G_c^\downarrow|$. Query times are reported for different time limits (given in minutes) The variant with the least absolute (relative) deviations is highlighted in dark (light) gray.

| | | # CELLS | | | | | | | |
|---|---|---|---|---|---|---|---|---|---|
| $|G_c^\downarrow|$ | $\tau$ | 128 | 256 | 512 | 1 024 | 2 048 | 4 096 | 8 192 | 16 384 |
| 128 | 10 | 4.38 | 2.92 | 2.02 | 1.53 | 1.09 | 1.05 | 1.29 | 2.18 |
| | 100 | 4.65 | 3.54 | 2.91 | 2.65 | 2.62 | 3.79 | 4.89 | 7.68 |
| | 500 | 10.42 | 8.67 | 8.05 | 7.67 | 8.12 | 10.78 | 14.43 | 21.29 |
| 256 | 10 | 4.34 | 2.86 | 1.98 | 1.50 | 1.07 | 1.03 | 1.27 | 2.16 |
| | 100 | 4.67 | 3.49 | 2.85 | 2.56 | 2.49 | 3.14 | 4.49 | 7.04 |
| | 500 | 10.31 | 8.50 | 7.79 | 7.56 | 7.60 | 9.29 | 13.02 | 18.69 |
| 512 | 10 | 4.32 | 2.82 | 1.92 | 1.80 | 0.94 | 1.01 | 1.24 | 2.13 |
| | 100 | 5.19 | 3.40 | 2.74 | 2.53 | 2.26 | 2.77 | 3.97 | 6.14 |
| | 500 | 10.57 | 8.21 | 7.33 | 6.70 | 6.69 | 7.82 | 11.28 | 15.27 |
| 1 024 | 10 | 4.27 | 2.73 | 1.84 | 1.39 | **0.88** | 0.97 | 1.24 | 2.10 |
| | 100 | 4.51 | 3.30 | 2.62 | 2.24 | 2.01 | 2.34 | 3.08 | 4.92 |
| | 500 | 9.90 | 7.82 | 6.78 | 5.90 | 5.65 | 6.07 | 7.56 | 10.70 |
| 2 048 | 10 | 4.29 | 2.71 | 1.82 | 1.04 | 0.95 | 0.89 | 1.28 | 2.17 |
| | 100 | 4.48 | 3.25 | 2.53 | 1.85 | 1.82 | 1.98 | 2.50 | 3.83 |
| | 500 | 9.71 | 7.56 | 6.37 | 5.12 | 4.81 | 4.86 | 5.49 | 7.13 |
| 4 096 | 10 | 4.26 | 2.75 | 1.87 | 1.07 | 1.01 | 0.89 | 1.32 | 2.25 |
| | 100 | 4.49 | 3.26 | 2.52 | 1.82 | **1.73** | 1.80 | 2.20 | 3.34 |
| | 500 | 9.64 | 7.48 | 6.14 | 4.84 | 4.34 | 4.17 | 4.38 | 5.51 |
| 8 192 | 10 | 4.42 | 3.05 | 1.96 | 1.19 | 1.13 | 1.14 | 1.41 | 2.37 |
| | 100 | 4.60 | 3.82 | 2.60 | 1.88 | 1.75 | 1.79 | 2.10 | 3.12 |
| | 500 | 9.65 | 7.94 | 6.09 | 4.75 | 4.14 | 3.81 | 3.84 | 4.64 |
| 16 384 | 10 | 4.54 | 3.01 | 2.17 | 1.45 | 1.49 | 1.46 | 1.65 | 2.61 |
| | 100 | 4.78 | 3.54 | 2.80 | 2.07 | 1.93 | 1.94 | 2.21 | 3.18 |
| | 500 | 9.81 | 7.58 | 6.23 | 4.86 | 4.20 | 3.80 | **3.73** | 4.37 |
| 32 768 | 10 | 4.88 | 3.41 | 2.57 | 2.07 | 2.14 | 2.14 | 2.11 | 2.99 |
| | 100 | 5.16 | 3.94 | 3.19 | 2.52 | 2.39 | 2.36 | 2.57 | 3.49 |
| | 500 | 10.15 | 7.94 | 6.57 | 5.20 | 4.53 | 4.07 | 3.97 | 4.55 |
| 65 536 | 10 | 5.55 | 4.10 | 3.44 | 3.15 | 3.18 | 3.15 | 3.31 | 3.73 |
| | 100 | 5.91 | 4.69 | 4.02 | 3.43 | 3.32 | 3.30 | 3.42 | 4.18 |
| | 500 | 10.89 | 8.67 | 7.32 | 5.94 | 5.24 | 4.78 | 4.65 | 5.20 |

Table 9. Sequential queries of isoPHAST

| | PREPRO [s] | | | | QUERIES | | |
|---|---:|---:|---:|---:|---:|---:|---:|
| algorithm | contract cells | ecc./ diam. | contract core/graph | comp. dist. tbl. | upward time [ms] | scanning time [ms] | # active cells |
| isoPHAST-CD | 17.37 | 0.58 | – | – | 15.76 | 15.87 | 329 |
| isoPHAST-CP | 19.43 | 0.74 | 1 171.57 | – | 1.20 | 7.83/21.97 | 221 |
| isoPHAST-DT | 13.41 | 5.26 | 672.19 | 307.94 | 0.38 | 24.37 | 1 439 |

Table 10. Parallel queries of isoPHAST

| | PREPRO [s] | | | | QUERIES | | |
|---|---:|---:|---:|---:|---:|---:|---:|
| algorithm | contract cells | ecc./ diam. | contract core/graph | comp. dist. tbl. | upward time [ms] | scanning time [ms] | # active cells |
| isoPHAST-CD | 29.51 | 1.49 | – | – | 4.47 | 3.76 | 64 |
| isoPHAST-CP | 29.49 | 0.95 | 1 392.66 | – | 1.04 | 2.13/4.68 | 77 |
| isoPHAST-DT | 17.76 | 3.00 | 717.72 | 75.08 | 0.40 | 3.38 | 587 |

**Detailed Timings for isoPHAST.** Tables 9 and 10 show details about sequential and parallel query times, respectively. We report duration of different phases of preprocessing and queries. Moreover, we show the average number of active cells.

**Scaling of Parallel Queries.** Finally, Figure 4 shows parallel query times (using 16 cores) for varying time limits for the same random queries as in Section 6. As mentioned before, curves mainly resemble the ones in the single-core setup. For the isoPHAST-CP, the linear sweep through the core graph is slightly faster now, since the core graph is smaller. The isoPHAST-CD variant is the best technique for mid-range queries again, however, as before, query performance get worse for larger time limits. Note that the performance gap between isoCRP and isoGRASP is smaller when using multiple cores (as explained in the main text, we conjecture that isoGRASP is limited by the memory bandwidth).

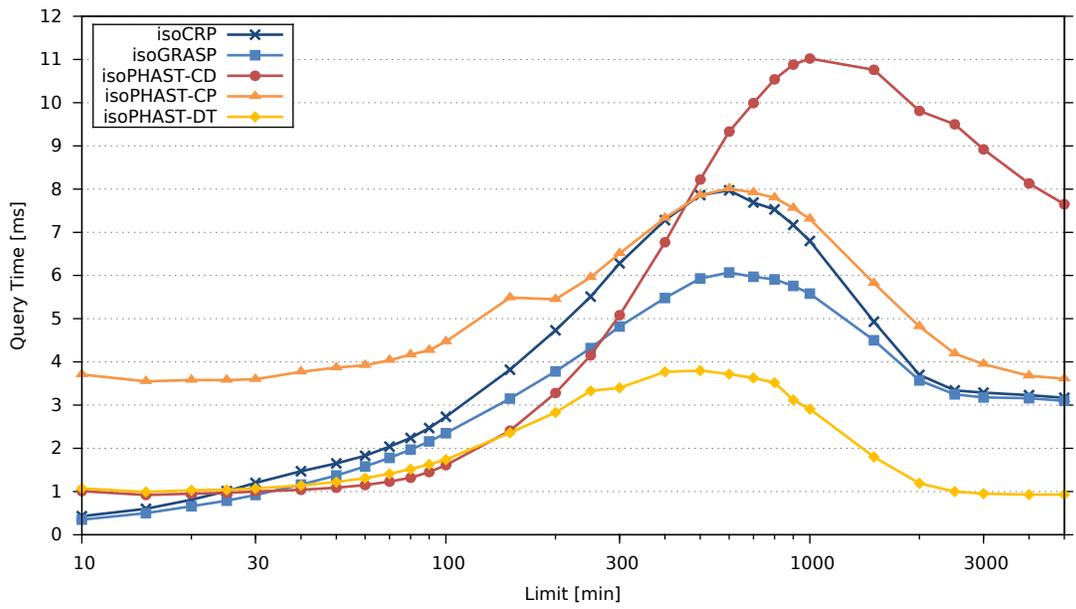

**Figure 4.** Parallel query times using 16 threads for varying time limits.